\documentclass{article} 
\usepackage{iclr2023_conference,times}

\usepackage{amsmath,amsfonts,bm}

\def\eqref#1{equation~\ref{#1}}

\def\1{\bm{1}}

\def\rvtheta{{\boldsymbol{\theta}}}

\def\rvs{{\mathbf{s}}}

\def\rvx{{\mathbf{x}}}
\def\rvy{{\mathbf{y}}}

\DeclareMathAlphabet{\mathsfit}{\encodingdefault}{\sfdefault}{m}{sl}
\SetMathAlphabet{\mathsfit}{bold}{\encodingdefault}{\sfdefault}{bx}{n}

\newcommand{\bTh}{\boldsymbol{\Theta}}
\newcommand{\E}{\mathbb{E}}

\newcommand{\KL}{D_{\mathrm{KL}}}

\usepackage{graphicx}
\usepackage{hyperref}
\usepackage{url}
\usepackage{xspace}
\usepackage{wrapfig}

\title{Bayesian calibration of differentiable\\ agent-based models}

\author{Arnau Quera-Bofarull 
\\
Department of Computer Science\\
University of Oxford\\
\texttt{arnau.quera-bofarull@cs.ox.ac.uk} \\
\And
Ayush Chopra \\
Media Lab \\
Massachusetts Institute of Technology \\
\texttt{ayushc@mit.edu} \\
\AND
Anisoara Calinescu
\\
Department of Computer Science\\
University of Oxford\\
\texttt{ani.calinescu@cs.ox.ac.uk} \\
\And
Michael Wooldridge
\\
Department of Computer Science\\
University of Oxford\\
\texttt{mjw@cs.ox.ac.uk} \\
\AND
Joel Dyer \\
Department of Computer Science \& Institute for New Economic Thinking \\
University of Oxford \\
\texttt{joel.dyer@cs.ox.ac.uk}
}

\newcommand{\abm}{\textsc{abm}}
\newcommand{\abc}{\textsc{abc}}
\newcommand{\gradabm}{\textsc{Gradabm}\xspace}
\newcommand{\gradjune}{\textsc{Gradabm-June}\xspace}
\newcommand{\june}{\textsc{June}\xspace}
\newcommand{\sbi}{\textsc{sbi}}
\newcommand{\gvi}{\textsc{gvi}}

\iclrfinalcopy 
\begin{document}

\maketitle

\begin{abstract}
Agent-based modelling ({\abm}ing) is a powerful and intuitive approach to modelling complex systems; however, the intractability of {\abm}s' likelihood functions and the non-differentiability of the mathematical operations comprising these models present a challenge to their use in the real world. These difficulties have in turn generated research on approximate Bayesian inference methods for {\abm}s and on constructing differentiable approximations to arbitrary {\abm}s, but little work has been directed towards designing approximate Bayesian inference techniques for the specific case of differentiable {\abm}s. In this work, we aim to address this gap and discuss how generalised variational inference procedures may be employed to provide misspecification-robust Bayesian parameter inferences for differentiable {\abm}s. We demonstrate with experiments on a differentiable {\abm} of the COVID-19 pandemic that our approach can result in accurate inferences, and discuss 
avenues for future work. 
\end{abstract}

\section{Introduction}

Agent-based models ({\abm}s) are growing in popularity as a modelling paradigm for complex systems in various fields, such as economics \citep{Baptista2016, paulin2019understanding} and epidemiology \citep{aylett2021june}. Such models simulate the interactions and decisions of a set of autonomous entities, where the rules governing those decisions and interactions are often nonlinear and stochastic. 

While this modelling approach provides considerable flexibility to the modeller, the complex structure and stochastic nature of many {\abm}s raise two key difficulties in deploying them in practice:
\begin{itemize}
    \item {\abm}s typically lack a tractable likelihood function -- denoted with $p(\rvx \mid \rvtheta)$, where $\rvx$ is the model output and $\rvtheta \in \bTh \subseteq \mathbb{R}^d$ are the $d$-dimensional free parameters of the model -- which complicates the problem of calibrating the model's free parameters;
    \item the mathematical expressions comprising the {\abm}'s specification are typically non-differentiable, 
    which presents a barrier to the use of gradient-based methods 
    in problems such as model calibration.
\end{itemize}

Over recent years, two active lines of research have emerged that seek to address each of these problems: a growing literature on approximate parameter inference techniques for {\abm}s, which has become increasingly focused on Bayesian methods \citep[see e.g.][]{grazzini2017bayesian, platt2021bayesian, dyer2022black}; and the development of techniques for building differentiable approximations to initially non-differentiable {\abm}s \citep{chopra2022differentiable, monti2022learning}. However, the combination of Bayesian parameter calibration methods and differentiable {\abm}s has 
not yet been considered. 

In this paper, we examine the problem of performing approximate Bayesian parameter inference for differentiable {\abm}s, and consider an approach that exploits the differentiability of the differentiable agent-based simulator. We discuss how the approach we consider -- which is derived from the literature on generalised Bayesian inference \citep[see e.g.][]{bissiri2016general, knoblauch2022optimization} -- may enjoy favourable robustness properties in comparison to alternative techniques, enabling the differentiable {\abm} to be applied more successfully in misspecified settings.

\section{Background}

\subsection{Simulation-based Bayesian inference for agent-based models}

Simulation-based inference (\sbi) algorithms are a set of procedures for performing parameter inference for simulation models, such as {\abm}s. 
Bayesian approaches to {\sbi} for {\abm}s have gained popularity over recent years with the use of approximate Bayesian computation (\abc) \citep{tavare1997inferring, pritchard1999population, beaumont2002approximate}, see e.g. \citet{grazzini2017bayesian, van2015calibration}. Broadly speaking, {\abc} targets an approximate posterior $q_{\abc}$ of the form
\begin{equation}
    q_{\abc}(\rvtheta \mid \rvs_{\rvy}) \propto \hat{p}_{\abc}(\rvs_{\rvy} \mid \rvtheta)\, \pi(\rvtheta),
\end{equation}
where $\rvs_{\rvx}$ is some summary statistic of $\rvx$, $\hat{p}_{\abc}(\rvs_{\rvy} \mid \rvtheta)$ is some approximation to the model's likelihood function, and $\pi$ is a prior density over $\bTh$. 
A variety of choices for $\hat{p}_{\abc}(\rvs_{\rvy} \mid \rvtheta)$ 
have been explored in the literature; see \citet{grazzini2017bayesian} and \citet{platt2021bayesian} for examples and \citet{dyer2022black} for a broader review.

Beyond \abc{}, \citet{dyer2022black} introduces and motivates the use of a class of more recently developed {\sbi} procedures. 
These approaches offer considerable benefits, such as a vastly reduced 
simulation burden compared to alternative techniques, and the ability to flexibly accommodate data of different kinds (e.g. dynamic graph data, see \citet{dyer2022calibrating}) -- but their performance can deteriorate in misspecified settings \citep{cannon2022investigating}, motivating research into methods for improving their robustness \citep{wardrobust, kelly2023misspecification}.

\subsection{Differentiable agent-based models}
Differentiable agent-based models ({\gradabm}s) are a recent class of tensorized {\abm}s that offer a number of benefits, such as compatibility with gradient-based learning via automatic differentiation, and the ability to simulate million-size populations in a few seconds on commodity hardware, integrate with deep neural networks, and ingest heterogeneous data sources. Recent prior work 
has demonstrated the utility of {\gradabm}s for scalable and fast forward simulations \citep{fastabm_wsc}, 
end-to-end calibration by coupling with neural networks
\citep{calibnn_aamas}, as well as efficient validation using gradient-based sensitivity analyses \citep{quera-bofarull2023a}. Direct access to simulator gradients in {\gradabm}s allows the modeller to leverage a statistical model for calibration without the need to build approximate surrogates. While this has shown to help integrate diverse data sources to enable more accurate calibration, prior work has only leveraged such gradients 
to generate point estimates of parameters. 
However, maximising the real-world utility of {\gradabm}s will also require uncertainty quantification during model calibration tasks; this motivates the use of Bayesian approaches for model calibration. 

\begin{figure}
\begin{center}
\includegraphics[width=0.7\linewidth]{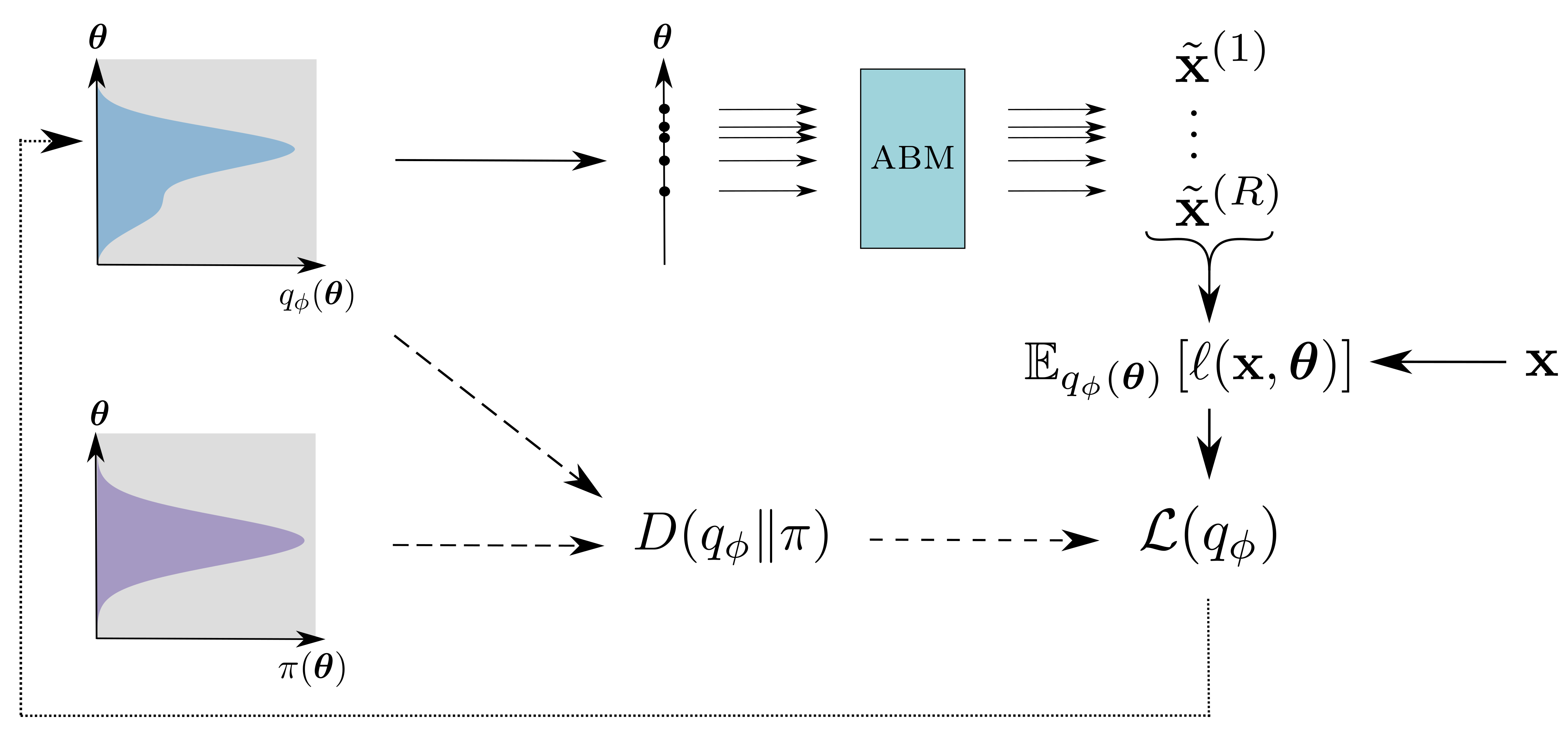}
\end{center}
\caption{A schematic of the method we employ. The posterior density estimator $q_{\phi}$ is encouraged to remain similar to the prior $\pi$ through some divergence $D$ (dashed arrows), while also being encouraged to generate simulations from the {\abm} that closely match the data $\rvx$ to which the model is being calibrated (solid arrows). The overall loss is a linear combination of these contributing terms and is used to inform the shape of $q_{\phi}$ (dotted arrow).}
\label{fig:method}
\end{figure}

\section{Method}
\label{sec:method}
 
To perform approximate Bayesian inference for a differentiable agent-based simulator, we consider a simulation-based Bayesian inference procedure with appealing robustness properties. In particular, we find a posterior density $q$ as the solution to a generalised variational inference (\gvi) problem, which has been shown to possess favourable robustness properties in comparison to methods that target the classical Bayesian posterior \citep{knoblauch2022optimization}. This enables us to obtain practically useful updates to the initial belief distribution, captured by the prior density $\pi$, when the agent-based model is not a perfect representation of the true data-generating mechanism (as is the case in reality).

G\textsc{vi} proceeds by first establishing a triple consisting of: a lower-bounded loss function/scoring rule $\ell : \mathcal{X} \times \bTh \to \mathbb{R}_{\geq 0}$ capturing some notion of discrepancy between the observed data $\rvx$ and the behaviour of the simulator at parameter value $\rvtheta$; a divergence $D$ between the posterior $q$ and prior $\pi$; and a search space $\mathcal{Q}$ of permitted solutions for $q$. 
Bayesian inference is then performed by 
solving the following optimisation problem:

\begin{equation}\label{eq:GVI}
    q^{*} = \arg\min_{q \in \mathcal{Q}} \mathcal{L}(q),\quad \quad \mathcal{L}(q) := \E_{q(\rvtheta)} \left[ \ell(\rvx, \rvtheta) \right] 
    + D\left(q \Vert \pi\right). 
\end{equation}
The solution $q^*$ is the generalised Bayesian posterior, 
while the \textit{classical} Bayesian posterior is obtained with 
$\ell(\rvx, \rvtheta) = - \log p(\rvx \mid \rvtheta)$ and $D$ as the Kullback-Liebler divergence $\KL$. 

To construct a flexible class of feasible solutions, we take $\mathcal{Q} = \{ q_{\phi} : \phi \in \Phi \}$, where $q_{\phi}$ is a normalising flow with trainable parameters $\phi$ taking values in some set $\Phi$. The 
flow parameters are then found as
\begin{equation}
    \phi^{*} =    \arg\min_{\phi \in \Phi} \Bigg\{ \E_{q_{\phi}(\rvtheta)} \left[ \ell(\rvx, \rvtheta) \right] 
    + D\left(q_{\phi} \Vert \pi \right) \Bigg\}.
\end{equation}
In this way -- owing to the ability to backpropagate pathwise gradients through the {\gradabm} -- we
can use the misspecification-robust inference framework of {\gvi} to calibrate the model parameters without recourse to potentially high-variance score-based gradient estimators \citep{mohamed2020monte}. 
A schematic of our method is shown in Figure \ref{fig:method}.

\section{Experiments}

We evaluate the proposed inference procedure by calibrating \gradjune \citep{gradabm_june_zenodo}, the differentiable version of the \june  model \citep{aylett2021june}. 
We use \gradjune to model the spread of SARS-CoV-2 in the London Borough of Camden, comprising approximately 250,000 people. 
For this experiment, we restrict ourselves to modelling the spread of infection in households, companies, and schools, 
each of which is regulated by the $\beta$-parameters that parameterize the spread of infection at different locations (see Appendix \ref{app:june} for details). 

We create a synthetic ground-truth time-series of SARS-CoV-2 infections for 30 days using the values $\beta_\mathrm{household} = 0.9$, $\beta_\mathrm{school} = 0.6$, and $\beta_\mathrm{company} = 0.3$,
%
\begin{wrapfigure}[25]{r}{0.48\textwidth}
      \vspace{-0.25cm}
  \centering
    \includegraphics[width=\linewidth]{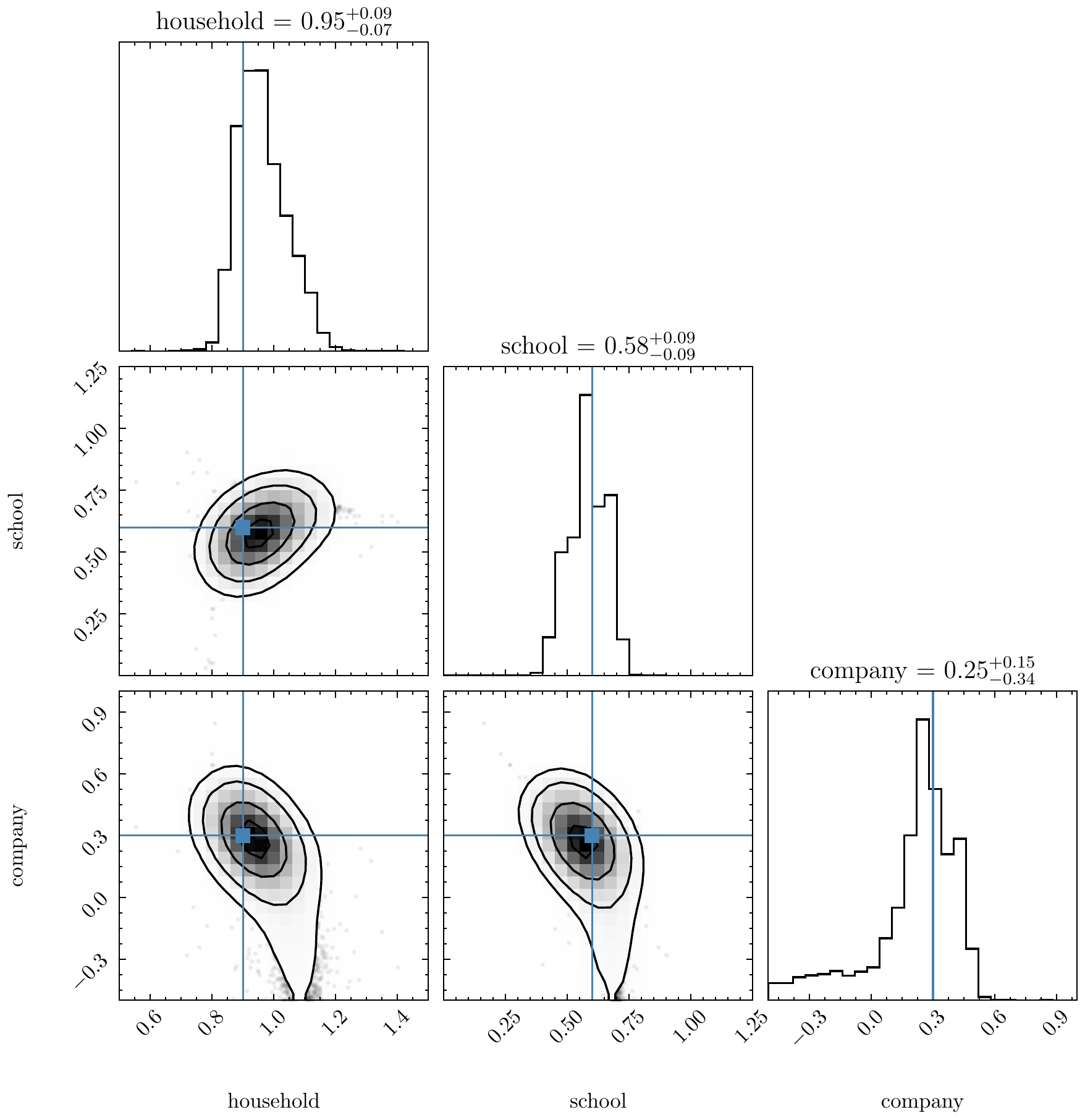}
    \caption{The inferred posterior distribution over the three calibrated $\beta$ parameters ($\beta_\mathrm{household}$, $\beta_\mathrm{school}$ and $\beta_\mathrm{company}$). The marginal densities are shown on the diagonal, and the off-diagonals show the bivariate joint densities for all pairs of parameters. The parameters that generated the pseudo-true synthetic dataset are shown with blue lines and points.}
    \label{fig:posterior}
\end{wrapfigure}
and obtain a posterior density by training a Neural Spline Flow (NSF) \citep{nsf}. We take
\begin{equation}\label{eq:our_ell}
    \ell\left(\rvx, \rvtheta\right) = \mathbb{E}_{\tilde{\rvx} \sim p(\cdot \mid \rvtheta)}\left[\sum_{t=1}^T \frac{\| \rvx_t - \tilde{\rvx}_t \|^2}{w}\right]
\end{equation}
in the loss function (\autoref{eq:GVI}), where $\rvx_t$ is the logarithm of the number of infections per time-step and the hyperparameter $w > 0$ balances the relative influence of the scoring rule to the divergence-measuring term. We further choose $D = \KL$ for simplicity. Details of the neural network architecture and additional training hyperparameters are provided in Appendix \ref{app:nn}.  

In this way, we obtain a posterior density over 
$\rvtheta$, which we show in \autoref{fig:posterior}. 
We see 
that the flow assigns high posterior density to the generating 
parameters, suggesting that the flow has assigned posterior mass to appropriate regions of $\bTh$. 
Of particular interest is the 2-dimensional projection of the density in the household-company and school-company plane, where we observe a trade-off between the contact intensity at households and companies. Indeed, it may be hard to distinguish where exactly infections are taking place when only the \textit{overall} number of cases is observed.

\begin{figure}[b]
    \centering
    \includegraphics[width=0.85\linewidth]{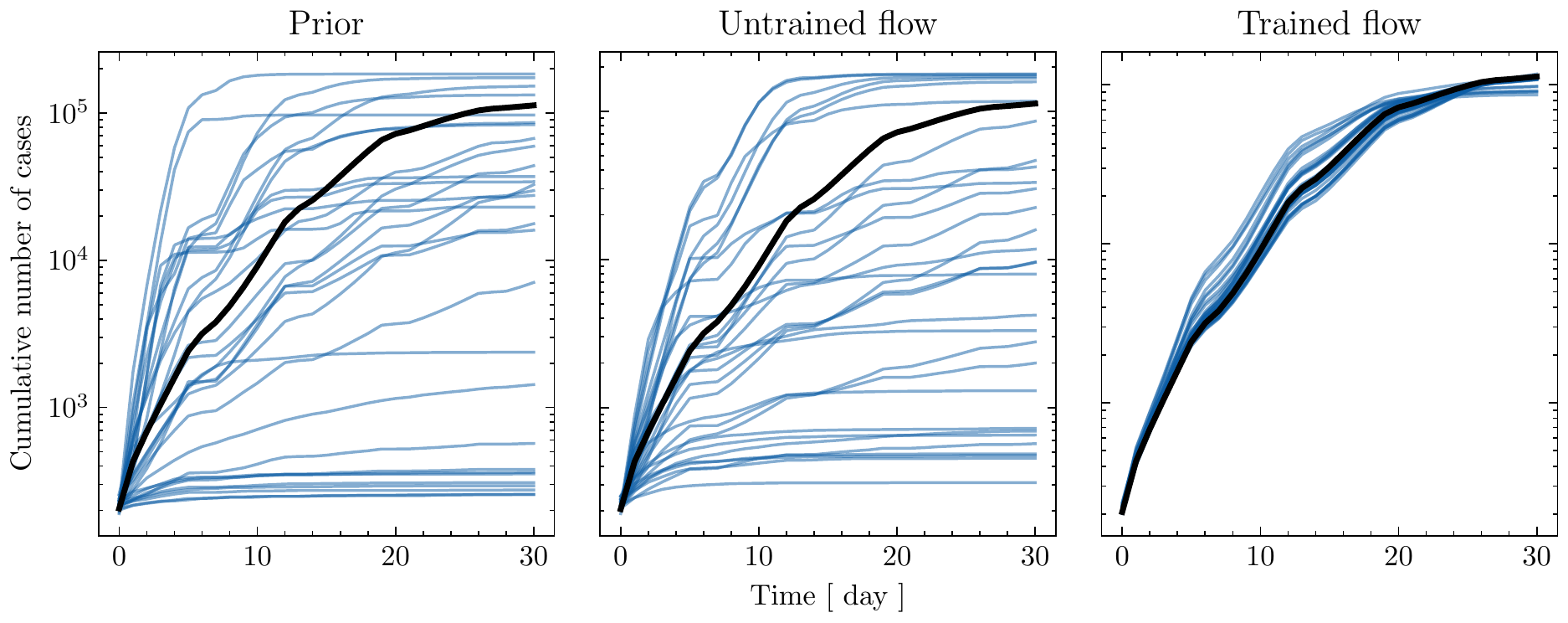}
    \caption{A comparison of the pseudo-true dataset (black curve) to simulations (blue curves) generated by parameters drawn from the prior density (\textbf{left}), the untrained normalising flow (\textbf{middle}), and the trained normalising flow (\textbf{right}).}
    \label{fig:comparison}
\end{figure}

We also show in Figure \ref{fig:comparison} a comparison between the pseudo-true dataset used to calibrate the model and simulations from the {\abm} generated by parameters drawn from the prior, the untrained flow, and the trained flow. From this, we see that the trained flow generates simulations that much better match the pseudo-true data than simulations generated by the prior or untrained flows, suggesting that our inference scheme has been successful in this model calibration task. Overall, the number of simulations required to train the flow was 2,500, which is small in comparison to e.g. {\abc}.

\section{Discussion \& conclusion}


In this paper, we consider how the task of Bayesian parameter calibration may be performed for differentiable {\abm}s. We discuss how the ability to backpropagate gradients through the agent-based simulator in a pathwise manner provides us with immediate access to a class of Bayesian inference methods known as generalised variational inference, and propose an approach drawn from this class of methods due to the fact that they may remedy misspecification-related problems more readily than existing approximate Bayesian inference methods for {\abm}s. Through experiments with \gradjune, a differentiable {\abm} of the COVID-19 pandemic in England, we demonstrate that our approach can provide accurate Bayesian inferences. We aim to develop this work into a full paper, at which point we will release the code for reproducing these results.

In future, we will test this method on real-world data and compare its performance against alternative {\sbi} techniques. We will also extend this work to the case of multiple observed \emph{iid} datasets $\rvx$ from some real-world density $p(\rvx)$ by considering the case of a conditional density estimator (e.g. a conditional normalising flow) and by training instead on the following loss function: 
\begin{equation}\label{eq:iid_gvi}
    q^* = \arg\min_{q \in \mathcal{Q}} \Bigg\{ \E_{q(\rvtheta \mid \rvx) p(\rvx)} \left[ \ell(\rvx, \rvtheta) \right] 
    + \E_{p(\rvx)}\left[D\left(q(\cdot \mid \rvx) \Vert \pi(\cdot) \right)\right] \Bigg\}.
\end{equation}
Once again, the choice $\ell(\rvx, \rvtheta) = -\log p(\rvx \mid \rvtheta)$ and $D = \KL$ will yield classical Bayesian posteriors, while other choices generate generalised posteriors. This may enable us to deploy the same conditional density estimator and {\abm} over a variety of scenarios covered by the density $p(\rvx)$, without retraining. 

\bibliography{iclr2023_conference}

\begin{thebibliography}{27}
\providecommand{\natexlab}[1]{#1}
\providecommand{\url}[1]{\texttt{#1}}
\expandafter\ifx\csname urlstyle\endcsname\relax
  \providecommand{\doi}[1]{doi: #1}\else
  \providecommand{\doi}{doi: \begingroup \urlstyle{rm}\Url}\fi

\bibitem[Aylett-Bullock et~al.(2021)Aylett-Bullock, Cuesta-Lazaro,
  Quera-Bofarull, Icaza-Lizaola, Sedgewick, Truong, Curran, Elliott, Caulfield,
  Fong, et~al.]{aylett2021june}
Joseph Aylett-Bullock, Carolina Cuesta-Lazaro, Arnau Quera-Bofarull, Miguel
  Icaza-Lizaola, Aidan Sedgewick, Henry Truong, Aoife Curran, Edward Elliott,
  Tristan Caulfield, Kevin Fong, et~al.
\newblock June: open-source individual-based epidemiology simulation.
\newblock \emph{Royal Society open science}, 8\penalty0 (7):\penalty0 210506,
  2021.

\bibitem[{Aylett-Bullock} et~al.(2021){Aylett-Bullock}, {Cuesta-Lazaro},
  {Quera-Bofarull}, Katta, Pham, Hoover, Strobelt, Jimenez, Sedgewick, Evers,
  Kennedy, Harlass, Maina, Hussien, and
  {Luengo-Oroz}]{aylett-bullockOperationalResponseSimulation2021}
Joseph {Aylett-Bullock}, Carolina {Cuesta-Lazaro}, Arnau {Quera-Bofarull},
  Anjali Katta, Katherine~Hoffmann Pham, Benjamin Hoover, Hendrik Strobelt,
  Rebeca~Moreno Jimenez, Aidan Sedgewick, Egmond~Samir Evers, David Kennedy,
  Sandra Harlass, Allen Gidraf~Kahindo Maina, Ahmad Hussien, and Miguel
  {Luengo-Oroz}.
\newblock Operational response simulation tool for epidemics within refugee and
  {{IDP}} settlements: {{A}} scenario-based case study of the {{Cox}}'s
  {{Bazar}} settlement.
\newblock \emph{PLOS Computational Biology}, 17\penalty0 (10):\penalty0
  e1009360, October 2021.
\newblock ISSN 1553-7358.
\newblock \doi{10.1371/journal.pcbi.1009360}.

\bibitem[Baptista et~al.(2016)Baptista, Farmer, Hinterschweiger, Low, Tang, and
  Uluc]{Baptista2016}
Rafa Baptista, J~Doyne Farmer, Marc Hinterschweiger, Katie Low, Daniel Tang,
  and Arzu Uluc.
\newblock {Macroprudential policy in an agent-based model of the UK housing
  market}.
\newblock \emph{Bank of England Working Paper}, 2016.

\bibitem[Beaumont et~al.(2002)Beaumont, Zhang, and
  Balding]{beaumont2002approximate}
Mark~A Beaumont, Wenyang Zhang, and David~J Balding.
\newblock {Approximate Bayesian computation in population genetics}.
\newblock \emph{Genetics}, 162\penalty0 (4):\penalty0 2025--2035, 2002.

\bibitem[Bissiri et~al.(2016)Bissiri, Holmes, and Walker]{bissiri2016general}
Pier~Giovanni Bissiri, Chris Holmes, and Stephen~G Walker.
\newblock A general framework for updating belief distributions.
\newblock \emph{Journal of the Royal Statistical Society: Series B (Statistical
  Methodology)}, 78\penalty0 (5):\penalty0 1103, 2016.

\bibitem[Cannon et~al.(2022)Cannon, Ward, and Schmon]{cannon2022investigating}
Patrick Cannon, Daniel Ward, and Sebastian~M Schmon.
\newblock Investigating the impact of model misspecification in neural
  simulation-based inference.
\newblock \emph{arXiv preprint arXiv:2209.01845}, 2022.

\bibitem[Chopra et~al.(2021)Chopra, Raskar, Subramanian, Krishnamurthy, Gel,
  Romero-Brufau, Pasupathy, and Kingsley]{fastabm_wsc}
Ayush Chopra, Ramesh Raskar, Jayakumar Subramanian, Balaji Krishnamurthy,
  Esma~S Gel, Santiago Romero-Brufau, Kalyan~S Pasupathy, and Thomas~C
  Kingsley.
\newblock Deepabm: scalable and efficient agent-based simulations via geometric
  learning frameworks-a case study for covid-19 spread and interventions.
\newblock In \emph{2021 Winter Simulation Conference (WSC)}, pp.\  1--12. IEEE,
  2021.

\bibitem[Chopra et~al.(2022{\natexlab{a}})Chopra, Rodriguez, Subramanian,
  Krishnamurthy, Prakash, and Raskar]{calibnn_aamas}
Ayush Chopra, Alexander Rodriguez, Jayakumar Subramanian, Balaji Krishnamurthy,
  B~Aditya Prakash, and Ramesh Raskar.
\newblock {Differentiable Agent-based Epidemiology}.
\newblock \emph{AAMAS 2023}, 2022{\natexlab{a}}.

\bibitem[Chopra et~al.(2022{\natexlab{b}})Chopra, Rodr{\'\i}guez, Subramanian,
  Krishnamurthy, Prakash, and Raskar]{chopra2022differentiable}
Ayush Chopra, Alexander Rodr{\'\i}guez, Jayakumar Subramanian, Balaji
  Krishnamurthy, B~Aditya Prakash, and Ramesh Raskar.
\newblock {Differentiable Agent-based Epidemiology}.
\newblock \emph{arXiv preprint arXiv:2207.09714}, 2022{\natexlab{b}}.

\bibitem[Durkan et~al.(2019)Durkan, Bekasov, Murray, and Papamakarios]{nsf}
Conor Durkan, Artur Bekasov, Iain Murray, and George Papamakarios.
\newblock Neural spline flows.
\newblock In H.~Wallach, H.~Larochelle, A.~Beygelzimer, F.~d\textquotesingle
  Alch\'{e}-Buc, E.~Fox, and R.~Garnett (eds.), \emph{Advances in Neural
  Information Processing Systems}, volume~32. Curran Associates, Inc., 2019.
\newblock URL
  \url{https://proceedings.neurips.cc/paper/2019/file/7ac71d433f282034e088473244df8c02-Paper.pdf}.

\bibitem[Dyer et~al.(2022{\natexlab{a}})Dyer, Cannon, Farmer, and
  Schmon]{dyer2022black}
Joel Dyer, Patrick Cannon, J~Doyne Farmer, and Sebastian Schmon.
\newblock {Black-box Bayesian inference for economic agent-based models}.
\newblock \emph{arXiv preprint arXiv:2202.00625}, 2022{\natexlab{a}}.

\bibitem[Dyer et~al.(2022{\natexlab{b}})Dyer, Cannon, Farmer, and
  Schmon]{dyer2022calibrating}
Joel Dyer, Patrick Cannon, J~Doyne Farmer, and Sebastian~M Schmon.
\newblock Calibrating agent-based models to microdata with graph neural
  networks.
\newblock \emph{arXiv preprint arXiv:2206.07570}, 2022{\natexlab{b}}.

\bibitem[Grazzini et~al.(2017)Grazzini, Richiardi, and
  Tsionas]{grazzini2017bayesian}
Jakob Grazzini, Matteo~G Richiardi, and Mike Tsionas.
\newblock Bayesian estimation of agent-based models.
\newblock \emph{Journal of Economic Dynamics and Control}, 77:\penalty0 26--47,
  2017.

\bibitem[Kelly et~al.(2023)Kelly, Nott, Frazier, Warne, and
  Drovandi]{kelly2023misspecification}
Ryan~P Kelly, David~J Nott, David~T Frazier, David~J Warne, and Chris Drovandi.
\newblock {Misspecification-robust Sequential Neural Likelihood}.
\newblock \emph{arXiv preprint arXiv:2301.13368}, 2023.

\bibitem[Knoblauch et~al.(2022)Knoblauch, Jewson, and
  Damoulas]{knoblauch2022optimization}
Jeremias Knoblauch, Jack Jewson, and Theodoros Damoulas.
\newblock An optimization-centric view on bayes’ rule: Reviewing and
  generalizing variational inference.
\newblock \emph{Journal of Machine Learning Research}, 23\penalty0
  (132):\penalty0 1--109, 2022.

\bibitem[Mohamed et~al.(2020)Mohamed, Rosca, Figurnov, and
  Mnih]{mohamed2020monte}
Shakir Mohamed, Mihaela Rosca, Michael Figurnov, and Andriy Mnih.
\newblock Monte carlo gradient estimation in machine learning.
\newblock \emph{The Journal of Machine Learning Research}, 21\penalty0
  (1):\penalty0 5183--5244, 2020.

\bibitem[Monti et~al.(2022)Monti, Pangallo, Morales, and
  Bonchi]{monti2022learning}
Corrado Monti, Marco Pangallo, Gianmarco De~Francisci Morales, and Francesco
  Bonchi.
\newblock On learning agent-based models from data.
\newblock \emph{arXiv preprint arXiv:2205.05052}, 2022.

\bibitem[Paulin et~al.(2019)Paulin, Calinescu, and
  Wooldridge]{paulin2019understanding}
James Paulin, Anisoara Calinescu, and Michael Wooldridge.
\newblock Understanding flash crash contagion and systemic risk: A micro--macro
  agent-based approach.
\newblock \emph{Journal of Economic Dynamics and Control}, 100:\penalty0
  200--229, 2019.

\bibitem[Platt(2021)]{platt2021bayesian}
Donovan Platt.
\newblock Bayesian estimation of economic simulation models using neural
  networks.
\newblock \emph{Computational Economics}, pp.\  1--52, 2021.

\bibitem[Pritchard et~al.(1999)Pritchard, Seielstad, Perez-Lezaun, and
  Feldman]{pritchard1999population}
Jonathan~K Pritchard, Mark~T Seielstad, Anna Perez-Lezaun, and Marcus~W
  Feldman.
\newblock Population growth of human {Y} chromosomes: a study of {Y} chromosome
  microsatellites.
\newblock \emph{Molecular biology and evolution}, 16\penalty0 (12):\penalty0
  1791--1798, 1999.

\bibitem[Quera-Bofarull et~al.(2023)Quera-Bofarull, Chopra, Aylett-Bullock,
  Cuesta-Lazaro, Calinescu, Raskar, and Wooldridge]{quera-bofarull2023a}
A~Quera-Bofarull, A~Chopra, J~Aylett-Bullock, C~Cuesta-Lazaro, A~Calinescu,
  R~Raskar, and M~Wooldridge.
\newblock Don’t simulate twice: one-shot sensitivity analyses via automatic
  differentiation.
\newblock Association for Computing Machinery, 2023.

\bibitem[Quera-Bofarull(2023)]{gradabm_june_zenodo}
Arnau Quera-Bofarull.
\newblock arnauqb/gradabm-june: Aamas zenodo release, February 2023.
\newblock URL \url{https://doi.org/10.5281/zenodo.7623959}.

\bibitem[Rozet \& Divo(2023)Rozet and Divo]{francois_rozet_2023_7625673}
François Rozet and Felix Divo.
\newblock francois-rozet/zuko: Zuko 0.1.4, February 2023.
\newblock URL \url{https://doi.org/10.5281/zenodo.7625673}.

\bibitem[Tavar{\'e} et~al.(1997)Tavar{\'e}, Balding, Griffiths, and
  Donnelly]{tavare1997inferring}
Simon Tavar{\'e}, David~J Balding, Robert~C Griffiths, and Peter Donnelly.
\newblock Inferring coalescence times from dna sequence data.
\newblock \emph{Genetics}, 145\penalty0 (2):\penalty0 505--518, 1997.

\bibitem[van~der Vaart et~al.(2015)van~der Vaart, Beaumont, Johnston, and
  Sibly]{van2015calibration}
Elske van~der Vaart, Mark~A Beaumont, Alice~SA Johnston, and Richard~M Sibly.
\newblock {Calibration and evaluation of individual-based models using
  Approximate Bayesian Computation}.
\newblock \emph{Ecological Modelling}, 312:\penalty0 182--190, 2015.

\bibitem[Vernon et~al.(2022)Vernon, Owen, {Aylett-Bullock}, {Cuesta-Lazaro},
  Frawley, {Quera-Bofarull}, Sedgewick, Shi, Truong, Turner, Walker, Caulfield,
  Fong, and Krauss]{vernonBayesianEmulationHistory2022a}
I.~Vernon, J.~Owen, J.~{Aylett-Bullock}, C.~{Cuesta-Lazaro}, J.~Frawley,
  A.~{Quera-Bofarull}, A.~Sedgewick, D.~Shi, H.~Truong, M.~Turner, J.~Walker,
  T.~Caulfield, K.~Fong, and F.~Krauss.
\newblock Bayesian emulation and history matching of {{JUNE}}.
\newblock \emph{Philosophical Transactions of the Royal Society A:
  Mathematical, Physical and Engineering Sciences}, 380\penalty0
  (2233):\penalty0 20220039, October 2022.
\newblock \doi{10.1098/rsta.2022.0039}.

\bibitem[Ward et~al.(2022)Ward, Cannon, Beaumont, Fasiolo, and
  Schmon]{wardrobust}
Daniel Ward, Patrick Cannon, Mark Beaumont, Matteo Fasiolo, and Sebastian~M
  Schmon.
\newblock {Robust Neural Posterior Estimation and Statistical Model Criticism}.
\newblock In \emph{Advances in Neural Information Processing Systems}, 2022.

\end{thebibliography}
\bibliographystyle{iclr2023_conference}

\appendix
\section{Appendix}

\subsection{The JUNE model}
\label{app:june}
The \june model \citep{aylett2021june} is a one-to-one agent-based epidemiological model that uses the English census to create a synthetic population at a very high resolution. The model has been used in a wide variety of settings such as to study the first and second wave of SARS-CoV2 in England \citep{vernonBayesianEmulationHistory2022a} and to mitigate disease spread in refugee settlements \citep{aylett-bullockOperationalResponseSimulation2021}. The model has been ported to the \gradabm framework to allow for much faster performance and gradient-based calibration \citep{chopra2022differentiable, gradabm_june_zenodo}. \june has a wide variety of configurable parameters regarding disease transmission and progression, vaccination, and non-pharmaceutical interventions. Given a susceptible agent exposed to an infection at location $L$, the probability of that agent getting infected is given by 
\begin{equation}
    p = 1 - \exp{\left(-\psi_s \; \beta_L \; \Delta t\sum_{i\in g}\mathcal I_i(t) \right) }
\end{equation}
where $\psi_s$ is the inherent susceptibility to infection of the agent, the summation is over all contacts with infected individuals at the location, $\mathcal I_i(t)$ is the time-dependent infectious profile of each infected agent, $\Delta t$ is the duration of the interaction, and, finally, $\beta_L$ is a location-specific parameter that models the difference in the nature of interactions for each location. Since the parameters $\beta_L$ are not physical parameters that can be measured, they are typically calibrated using data on the number of cases or fatalities over time. In this work, we focus on calibrating the $\beta_L$ parameters for our three modelled locations: households, schools, and companies.

\subsection{Neural network architecture and training}\label{app:nn}

To obtain a posterior density, we use a Neural Spline Flow (NSF) using \textsc{Zuko}\citep{francois_rozet_2023_7625673} as $q_{\phi}$, consisting of 3 transformations, each parameterized by a 3-layer fully connected neural network with 128 units in every layer. We did not explore the impact of different flow models or parameterisations on the calibration results.  We train the NSF model following the procedure described in Section \ref{sec:method} until the validation loss consisting of the forecasting and regularisation term (see Equation \ref{eq:GVI}) does not significantly change across epochs. At each epoch, we estimate 
\begin{equation*}
\KL(q_{\phi} \Vert \pi) \approx \frac{1}{R} \sum_{r=1}^R\left( \log q_{\phi}(\rvtheta^{(r)}) - \log \pi(\rvtheta^{(r)})\right), 
\end{equation*}
where $\rvtheta^{(r)} \overset{iid}{\sim} q_{\phi}$ and $R=10^4$. Likewise, the forecast loss (scoring rule) is estimated by following \autoref{eq:our_ell} with $T=10$. We obtain the best results by setting a regularisation weight of $w = 10^{-2}$.

\end{document}